\documentclass{iau}
\usepackage{graphicx}

\usepackage{natbib}
\usepackage{adjustbox}
\usepackage{longtable}
\usepackage{multirow}
\usepackage{siunitx}
\usepackage{subfigure}
\usepackage[colorlinks,allcolors=blue]{hyperref}

\title[UVIT observations of GGCs] %% give here short title %%
{UVIT Observations of UV-Bright	Stars in four Galactic Globular Clusters}

\author[R. Kumar et al.]   %% give here short author list %%
{Ranjan Kumar$^1$,         %%  \thanks{Present address: Fluid Mech Inc., 24 The Street, Lagos, Nigeria.},
 Ananta C. Pradhan$^1$,
 M. Parthasarathy$^2$,
 Devendra K. Ojha$^3$,
 Abhisek Mohapatra$^1$
 \and Jayant Murthy$^2$ }

\affiliation{$^1$Dept. of Physics and Astronomy, National Institute of Technology, Rourkela - 769 008, India \\ email: {\tt ranjankmr488@gmail.com}
\\[\affilskip]
$^2$Indian Institute of Astrophysics, Bangalore - 560 034, India
\\[\affilskip]
$^3$Dept. of Astronomy and Astrophysics, Tata Institute of Fundamental Research (TIFR), Mumbai - 400 005, India   }

\pubyear{2019}
\volume{351}  %% insert here IAU Symposium No.
\jname{Star Clusters: From the Milky Way to the Early Universe}
\editors{A. Bragaglia, M.B. Davies, A. Sills \& E. Vesperini, eds.}

\begin{document}

\maketitle

\begin{abstract}
We have performed photometric analysis of four Galactic globular clusters (GGCs): NGC 4147, NGC 4590, NGC 5053 and NGC 7492 using far-UV and near-UV filters of the Ultraviolet Imaging Telescope (UVIT) on-board AstroSat. With the help of color-magnitude diagrams (CMDs), we have identified $\sim$ 150 blue horizontal branch stars (BHBs), and  $\sim $ 40 blue straggler stars (BSS) in the four GGCs. We study the temperature and radial distribution of BHBs and BSS for the four GGCs.
\end{abstract}

\firstsection % if your document starts with a section,

\section{Introduction}

 The ultraviolet (UV) light in old stellar populations of Galactic globular clusters (GGCs) is dominated by hot stars, such as blue horizontal branch stars (BHBs), blue hook stars (BHk), extreme blue horizontal branch stars (eBHBs), blue straggler stars (BSS), post asymptotic giant branch stars (pAGB) and hot sub-dwarfs (sdB, sdO) \citep{rood98}. UV emission is strongly dependent on the source temperature, therefore hot BHBs, among others are the strongest contributors to the far-UV flux of GGCs. Far-UV colors are well co-related with horizontal branch morphology in GGCs and have been studied vastly using Hubble Space Telescope (HST), Galaxy Evolution Explorer (GALEX) and Ultraviolet Imaging Telescope (UVIT) observations \citep[see][] {Schiavon12, Dalessandro2012, Lagioia15, Pioto2015, Milone2015, Subramaniam2017, Sahu2019}. GGCs in halo region of Milky Way are less crowded in UV and hence, their observation in UV regime allows one to probe and separate out the bright stars effectively. \citet{Schiavon12} have presented UV CMDs of 44 GGCs using GALEX observations and \citet{Pioto2015} have done the same for 58 GGCs from HST observations. The relative contributions of the various types of stars and the factors that might lead to larger or smaller populations of UV-Bright stars have remained an open question \citep{Greggio90, Dorman95, Lee2002, Ambika2004, Jasniewicz2004, Sohn2006, Dalessandro2012}. We present UV photometric study of four GGCs: NGC 4147, NGC 4590, NGC 5053 and NGC 7492 with the observations taken from UVIT on-board AstroSat \citep{Tandon2017}. 

\vspace{-0.25 cm}
\section{Observation and Data Reduction}

The UVIT on-board AstroSat is performing observations in FUV (130-180 nm) and NUV (200-300 nm) bands, each having five filters with resolutions better than $1.8''$ since its launch in 2015 \citep{Tandon2017}. We have observed four GGCs, NGC 4147, NGC 4590, NGC 5053 and NGC 7492, using both the FUV and NUV filters of UVIT. The observational details are given in \autoref{tab:observation}. We reduced the data and produced good quality images of the clusters using a customized software package CCDLAB \citep{Postma2017}. We performed the crowded field photometry on the images using DAOPHOT package \citep{Stetson1987} available in IRAF. We have done the interstellar extinction correction on observed sources using \citet{Cardeli1989} extinction law considering E(B-V) value for each cluster from \citet{Schlegel1998} extinction map (see \autoref{tab:parameter}). We considered sources with AB magnitude limit up to 22.5 and 23.0 in FUV and NUV, respectively. We cross-matched the UVIT observed sources with GAIA DR2 catalogue \citep{Gaia2018} to get proper motion (PMRA, PMDEC) of observed sources. We separated out the cluster members from the field stars by selecting $1 \sigma$ Gaussian distribution from mean PMRA and PMDEC as cluster members. The number of sources we obtained in FUV (NUV) as cluster members for GGCs, NGC 4590, NGC 5053, and NGC 7492 are 71(1875), 30(539) and 29(178), respectively. Since NGC 4147 has observation only with FUV filters, we could extract 45 cluster members in the FUV band.

        \begin{table}[t]
        \centering
        \caption{UVIT observation log of four GGCs}
        \label{tab:observation}
        \begin{adjustbox}{width=0.98\textwidth,height=0.11\textheight, keepaspectratio}
        \begin{tabular}{|c|c |c c c |c c c|}
         \hline
         \multicolumn{2}{|c|}{Telescope} &  & FUV &  & & NUV& \\ \hline
         \multicolumn{2}{|c|}{Filter Name}& BaF$_2$& Sapphire & Silica & NUVB15 & NUVB13 & NUVB4\\
         \multicolumn{2}{|c|}{$\lambda_{eff.}$ of filters (in $\si{\angstrom}$)} & 1541  & 1608 & 1717  & 2196  & 2447 & 2632\\ \hline
         Cluster Name & ($gl, gb$) in degrees & \multicolumn{6}{c|}{Exposure Time in seconds}\\ \hline
         NGC 4147 & ($252.85^\circ,77.19^\circ$) &  1536 & 1648 & 1209 & - & - & - \\ \hline
         NGC 4590 & ($299.62^\circ, 36.05^\circ$) & 860 & - & 1406 & 1221 & 1739 & 1607\\ \hline
         NGC 5053 & ($335.70^\circ,78.95^\circ$) & - & 1423 & 217 & - & 1503 & 817 \\ \hline 
         NGC 7492 & ($53.39^\circ,-63.48^\circ$) &  - & - & 850 & - & 3276 & 3014\\ \hline
   
        \end{tabular}
        \end{adjustbox}
        \end{table}
        
\vspace{-0.25 cm}
\section{Color Magnitude Diagram}

\begin{table}[b]
    \centering
    \caption{Cluster parameters of GGCs and extracted UV-Bright sources from CMDs.}
    \label{tab:parameter}
    
    \begin{adjustbox}{width={\textwidth},keepaspectratio}
    \begin{tabular}{|c|c|c|c|c|c||c|c|c|c|c|} \hline
     & \multicolumn{5}{c||}{Cluster parameters of GGCs}& \multicolumn{5}{c|}{Number of UV-Bright sources}\\ \hline
    GGCs  & Gal. lat. & Metallicity & Distance & E(B-V) & age & \multicolumn{2}{c|}{BHBs} & \multicolumn{2}{c|}{BSS} & eBHB  \\
         & in degree & [Fe/H]   & (kpc)  & (mag) &(Gyr) & FUV & NUV & FUV & NUV & \\ \hline
    NGC 7492 & $-63.48^\circ$ & -1.80 & 26.3 & 0.0310 & 12.0 & 28 & 28 & 5 & 5 & - \\ \hline
    NGC 5053 & $78.95^\circ$ & -2.27 & 17.4 & 0.0149 & 12.3 & 25 & 35 & 2 & 29 & 1 \\ \hline
    NGC 4590 & $36.05^\circ$ & -2.23 & 10.3 & 0.0523 & 12.5 & 57 & 97 & 3 & 7 & - \\ \hline
    NGC 4147 & $77.19^\circ$ & -1.80 & 19.3 & 0.0221  & 13.5 & 34 & - & 4 & - & - \\ \hline
    \end{tabular}
    \end{adjustbox}
\end{table}

\begin{figure*}
   \begin{center}
   \subfigure[ \label{fig:1a}]{\includegraphics[width=0.48\linewidth]{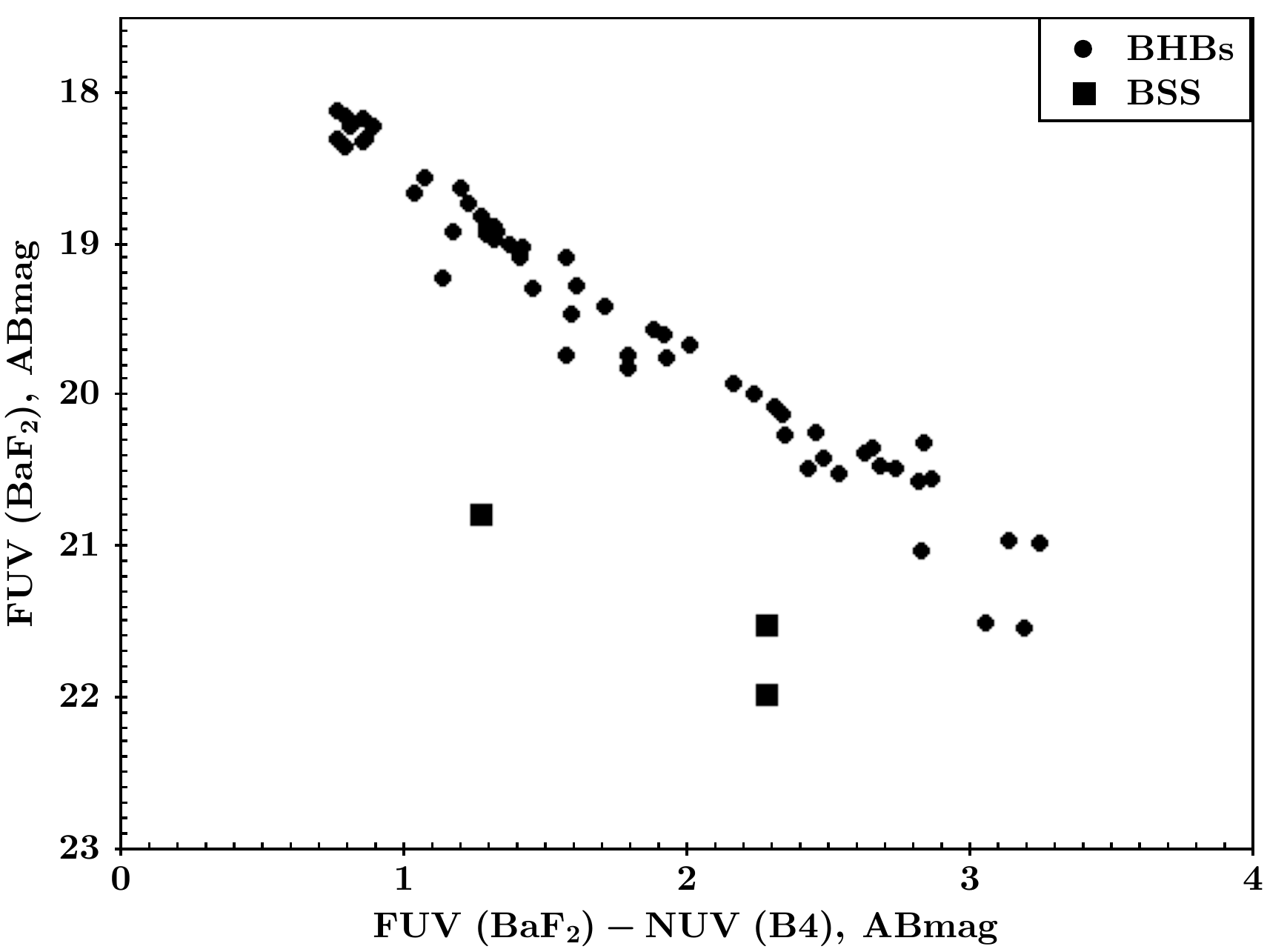}} 
   \subfigure[ \label{fig:1b}]{\includegraphics[width=0.48\linewidth]{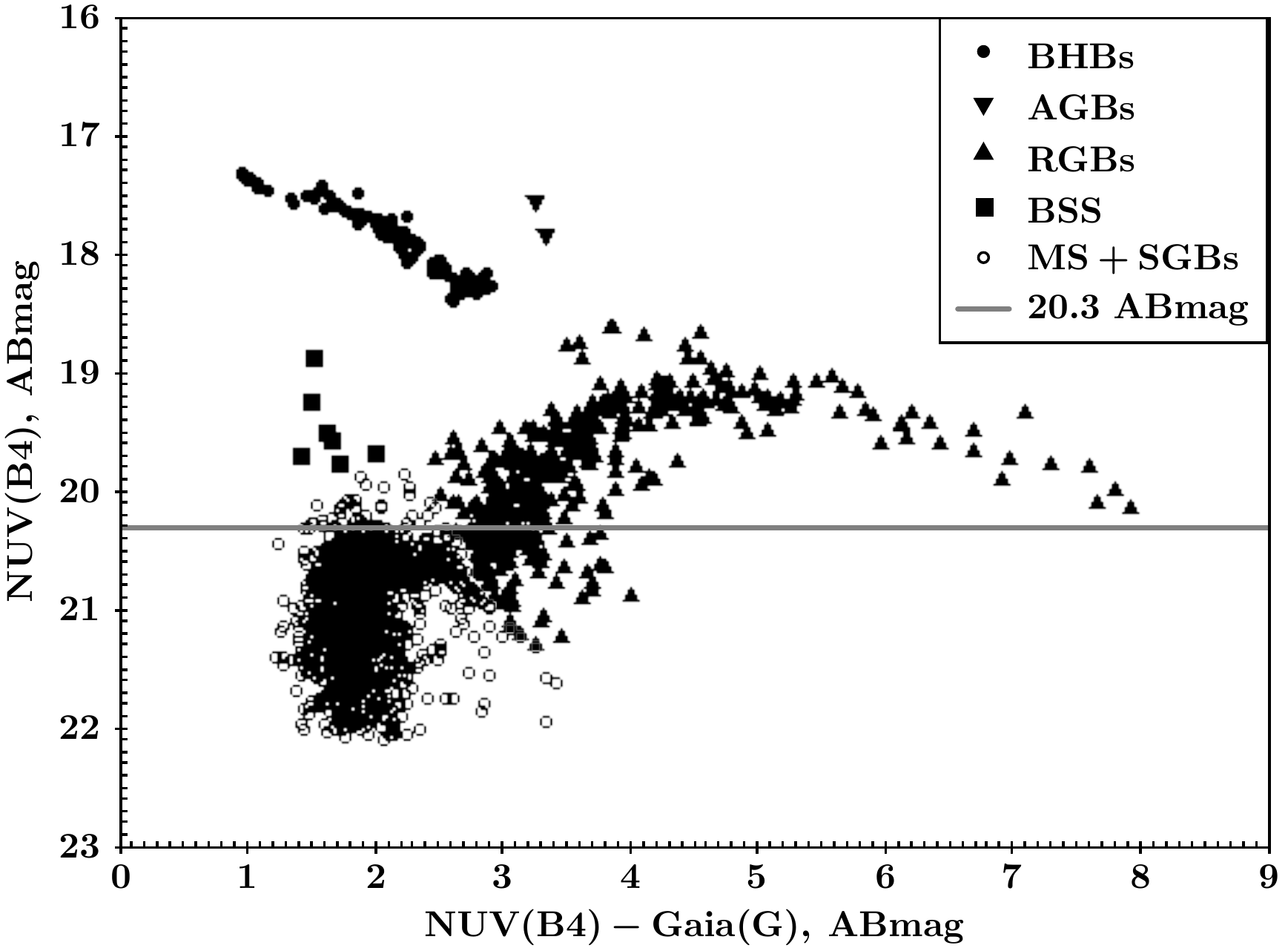}}
   \end{center}
    
  \caption{ Fig.~\ref{fig:1a} shows the FUV($BaF_2$) - NUV(B4) vs FUV($BaF_2$) and Fig.~\ref{fig:1b} shows the NUV(B4) - Gaia(G) vs NUV(B4) CMDs of NGC 4590. Various stellar populations present in the cluster are denoted with different symbols: filled circles show the BHBs, filled squares show BSS, lower triangles show AGBs, upper triangles show RGBs, and MS+SGBs are shown with open circles. Gray horizontal line in Fig.~\ref{fig:1b} shows the MS-turn off point in NUV(B4).}

  \label{fig:cmd}
\end{figure*}

After getting the catalog of sources for each cluster we have classified various stellar populations by plotting CMDs using various FUV, NUV and optical filters. We have shown the CMD, FUV($BaF_2$) - NUV(B4) vs FUV($BaF_2$) (Fig.~\ref{fig:1a}), and CMD, NUV(B4) - Gaia (G) vs NUV(B4) (Fig.~\ref{fig:1b}), for one of the clusters NGC 4590. The stellar populations present in the cluster were separated by identifying their various regions in the UV CMDs previously suggested by \citet{Schiavon12, Subramaniam2017, Sahu2019}. We can see that in FUV only BHBs and BSS are visible (Fig.~\ref{fig:1a}), whereas in NUV BHBs, BSS, RGBs, SGBs and MSs are clearly visible (Fig.~\ref{fig:1b}). We see the main sequence turn-off at 20.3 ABmag in NUV(B4) filter (Fig.~\ref{fig:1b}). We are able to see the MS branch in NUV band which is not visible in the FUV band.  We have seen a similar trend of distribution of sources in the CMDs of other clusters as well. A list of UV-Bright sources extracted for the four GGCs are given in \autoref{tab:parameter}.

\vspace{-0.25 cm}
\section{Temperature and Radial Distribution of BHBs and BSS}

\begin{figure}[b]
    \centering
    \includegraphics[totalheight={0.25\textheight},keepaspectratio]{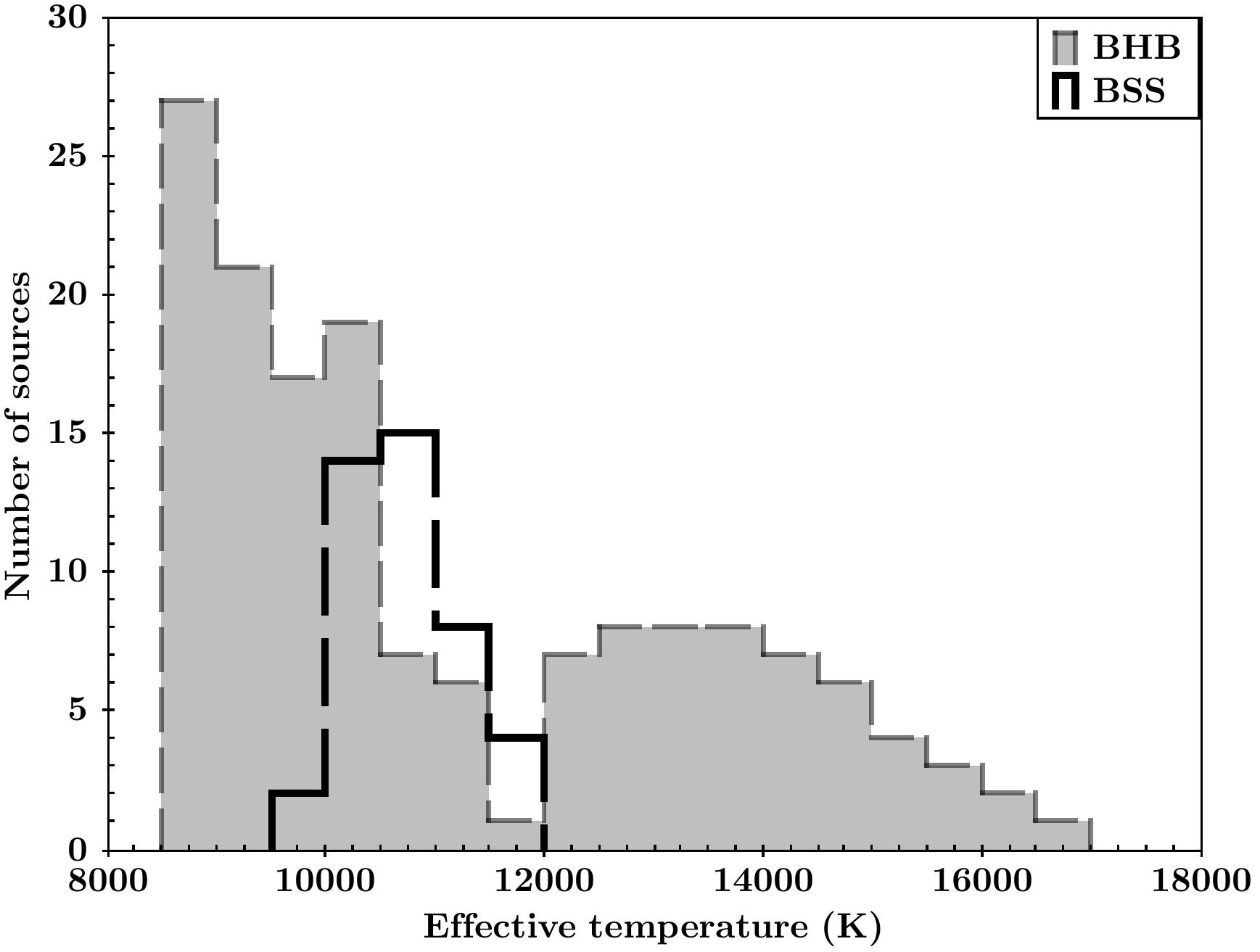}
    \caption{Temperature distribution of BHBs and BSS observed in the four GGCs.}
    \label{fig:teff}
\end{figure}

There have been several studies of temperature distribution of BHBs and BSS of GGCs in UV using GALEX \citep{Schiavon12}, HST \citep{Lagioia15} and UVIT \citep{Sahu2019} observations. \citet{Schiavon12, Lagioia15} have used zero age horizontal branch (ZAHB) models to extract effective temperature ($T_{eff}$) from the UV CMDs, whereas \citet{Sahu2019} have used the color-temperature relation from the Kurucz stellar atmosphere model \citep{Kurucz03} and the SED fitting on the BHBs and BSS sources. 
In order to extract out $T_{eff}$ of all the observed BHBs and BSS in our observed four GGCs, we have used the color-temperature relation from Kurucz stellar atmosphere model \citep{Kurucz03}. Based on the color combinations used in FUV and NUV CMDs for all the clusters, we generated theoretical colors for a range of temperature (4,000K to 30,000K) using cluster parameters of observed GGCs given in \autoref{tab:parameter} and a surface gravity, log(g)=4.0 for BHBs and BSS \citep[suggested by][for BHBs] {Lagioia15, Sahu2019}. Then observed colors and theoretical colors were matched, within $\sigma_{color}\leq0.01$ which have max $\Delta T = 100K$, to extract the corresponding $T_{eff}$ for each BHB star and BSS. The $T_{eff}$ distribution of 152 BHBs and 42 BSS observed with FUV and NUV filters of UVIT are plotted in Fig.~\ref{fig:teff}. The $T_{eff}$ of BHBs ranges from 8,500K to 17,000K, whereas the $T_{eff}$ of BSS ranges from 9,500K to 12,000K. The temperature distribution shows that there are two groups of BHBs present in the observed GGCs. We also see a gap between 11,500K and 12,000K, which suggests the Grundahl-jump in temperature distribution of BHBs \citep{Grundahl99}. 

We took the core radius, half light radius and the tidal radius from the updated catalogue of GGCs \citep{Harris2010} and then studied the radial distribution of cluster members observed in FUV and NUV bands. We found that the FUV emissions arise only from the BHBs and BSS sources which are concentrated within half light radius of the clusters. The NUV emissions arise from BHBs, BSS, RGBs and SGBs and their radial distribution goes up to the tidal radius of clusters. A detailed analysis on radial distribution and density of cluster members for all clusters will be done in forthcoming papers.

\vspace{10pt}
\textbf{Acknowledgements:} This research is supported by ISRO RESPOND PROJECT No. ISRO/RES/2/409/17-18. This publication uses data from the AstroSat mission of the Indian Space Research Organization (ISRO), archived at the Indian Space Science Data Centre (ISSDC).

\end{document}